\documentclass[runningheads]{llncs}

\usepackage[T1]{fontenc}
\usepackage{graphicx}
\usepackage{xcolor}
\usepackage{array}
\usepackage{tabularray}

\usepackage{comment}
\usepackage{subcaption}
\usepackage{multirow}

\def\rw{\bar{w}}
\def\rs{\bar{s}(t)}
\def\text{w}
\def\rtext{\rw}
\def\speech{s(t)}
\def\rspeech{\rs}
\def\rrspeech{\bar{\bar{s}}(t)}
\def\lengthregulator{$\mathcal{L}$}

\def\pair#1{Pair\# #1}
\def\subject#1{Sub\# #1}

\def\sub-words{{tokens}}
\def\sub-word{{token}}
\def\SP{{SentencePiece}}

\def\WaveNet{WaveGlow}
\def\eos{EoS}
\def\artts{{\tt Tacotron-2}}
\def\vits{{\tt VITS-TTS}}
\def\RevTxtFwdSpeech{{\tt rtfs-e2e-TTS}}
\def\ftts{{\color{black}{{\tt e2e-TTS}}}}
\def\rtts{{\color{black}{{\tt r-e2e-TTS}}}}

\begin{document}
\title{
Probing Human Articulatory Constraints in End-to-End TTS with Reverse and Mismatched Speech-Text Directions
}
\titlerunning{Reverse Speech TTS}
% If the paper title is too long for the running head, you can set
% an abbreviated paper title here
%
\author{Parth Khadse%\inst{1}%\orcidID{0000-1111-2222-3333} 
\and
Sunil Kumar Kopparapu%\inst{1} %\orcidID{1111-2222-3333-4444} 
% \and
% Third Author\inst{3}\orcidID{2222--3333-4444-5555}
}
\authorrunning{Parth Khadse, Sunil Kopparapu}
% First names are abbreviated in the running head.
% If there are more than two authors, 'et al.' is used.
%
\institute{TCS Research, Tata Consultancy Services Limited, Mumbai, India  
% \and Springer Heidelberg, Tiergartenstr. 17, 69121 Heidelberg, Germany
\email{\{parth.khadse,sunilkumar.kopparapu\}@tcs.com}\\
\url{http://www.tcs.com} 
% \and
% ABC Institute, Rupert-Karls-University Heidelberg, Heidelberg, Germany\\
% \email{\{abc,lncs\}@uni-heidelberg.de}
}

\maketitle              % typeset the header of the contribution
\begin{abstract}
% The abstract should briefly summarize the contents of the paper in
% 150--250 words.
    An end-to-end (e2e) text-to-speech (TTS) system is a deep architecture that learns to associate a text string with acoustic speech patterns from a curated dataset. It is expected that all aspects associated with speech production, such as phone duration, speaker characteristics, and intonation among other things are captured in the trained TTS model to enable the synthesized speech to be natural and intelligible. Human speech is complex, involving \textit{smooth} transitions between articulatory configurations (ACs). Due to anatomical constraints, some ACs are challenging to mimic or transition between. In this paper, we experimentally study if the constraints imposed by human anatomy have an implication on training an \ftts{} systems. We experiment with two \ftts{} architectures, namely, \artts{} an autoregressive model and \vits{} a non-autoregressive model. In this study, we build TTS systems using (a) forward text, forward speech (conventional, \ftts), (b) reverse text, reverse speech (\rtts), and (c) reverse text, forward speech (\RevTxtFwdSpeech). Experiments demonstrate that  \ftts{} systems are purely data-driven. Interestingly, the generated speech by {\rtts} systems exhibits better fidelity, better perceptual intelligibility, and better naturalness.

\keywords{TTS \and reverse speech \and end to end models \and tokenization}
% \keywords{First keyword  \and Second keyword \and Another keyword.}
\end{abstract}

\section{Introduction}
Articulation is the process of producing speech sounds by manipulating the air-stream from the lungs using the articulatory organs. The specific movements of the articulatory organs, called articulatory configuration (AC) determine the different speech sounds that humans produce. One could look at a sequence of AC smoothly transitioning from one AC to another resulting in spoken speech. 
Visible speech (VS) symbols \cite{bell1867},  introduced in 1867, provided visual representation of the position of the speech organs needed to articulate an individual sound. The concept of VS was used effectively
 to identify words that are difficult to articulate based on the degree of change in the position of articulators that is required to speak a word \cite{TRIPATHI2021101213}. 
While appreciating the fact that speech production is much more complex, namely, the next word or phrase that we produce is dependent on \textit{what} has been spoken prior not only linguistically but also in terms of \textit{how} we emote it in the form of intonation, there is also an aspect of breath cycle, pauses, filler words, stress on certain words, language grammar and much more. Aware of this,
we hypothesize that the language we speak has evolved over a period of time and the sounds that we produce are such that they are \textit{optimal} in terms of the amount of energy required to change the position of the articulators or transition from one AC to another when we speak a sentence. We call the speech produced with these constraints, to contain \textit{natural patterns} in spoken speech. Maybe, we as humans have refined or adapted the language structure, syntax, etc. to make it suitable, comfortable and easy for us to use. For example speaking ``txet'' (reverse of ``text'') is harder than ``text'' (see Appendix \ref{appendix:1}).

With growing need for voice user interfaces, there is a dire need for generating speech from text, also called text to speech (TTS). The area of TTS has evolved significantly from the rudimentary Klatt synthesizer \cite{Klatt1980} to current day \ftts{} models like \artts{} \cite{shen2018natural}, FastSpeech \cite{NEURIPS2019_f63f65b5}, Glow-TTS \cite{10.5555/3495724.3496400}, F5-TTS \cite{chen2025f5ttsfairytalerfakesfluent}, Variational Inference Text-to-Speech (\vits) \cite{kim2021conditional}. The \ftts{} systems are traditionally sequence to sequence (seq2seq) architectures that learn to map a text sequence and the corresponding audio. These \ftts{} are trained on a large amount of training text-speech corpora.  In this paper, we experimentally explore if these \ftts{} models benefit from  
% 
% This experiment might help explore if the model training benefits from
(a) the constraints imposed by the articulatory anatomy where only certain transitions of AC's are possible (see Appendix \ref{appendix:1}) or (b) some implicit patterns in the syntactic structure of the language and speech. We further assume that these patterns or AC transition constraints do not exist in a reverse audio. For clarity, if $s(t)$ is the spoken speech of duration $T$, namely, $0 \le t \le T$ then $s(T-t) = \rs$ is the reverse speech.
In this paper, we would like to study and understand if an \ftts{} architecture can be trained on an \textit{unnatural (to human) patterns} or in other words does the constraints on possible AC imposed by the articulatory anatomy or the smooth transition from one AC to another help the conventionally trained \ftts{} system. 

The rest of the paper is organized as follows. In Section \ref{sec:approach} we describe the approach that we adopted and in Section \ref{sec:experiments} we analyze the experiments that we conducted and conclude in Section \ref{sec:conclusion}.

\section{The Setup}
\label{sec:approach}

\begin{figure}[!htb]
    \centering
    \includegraphics[width=0.65\linewidth]{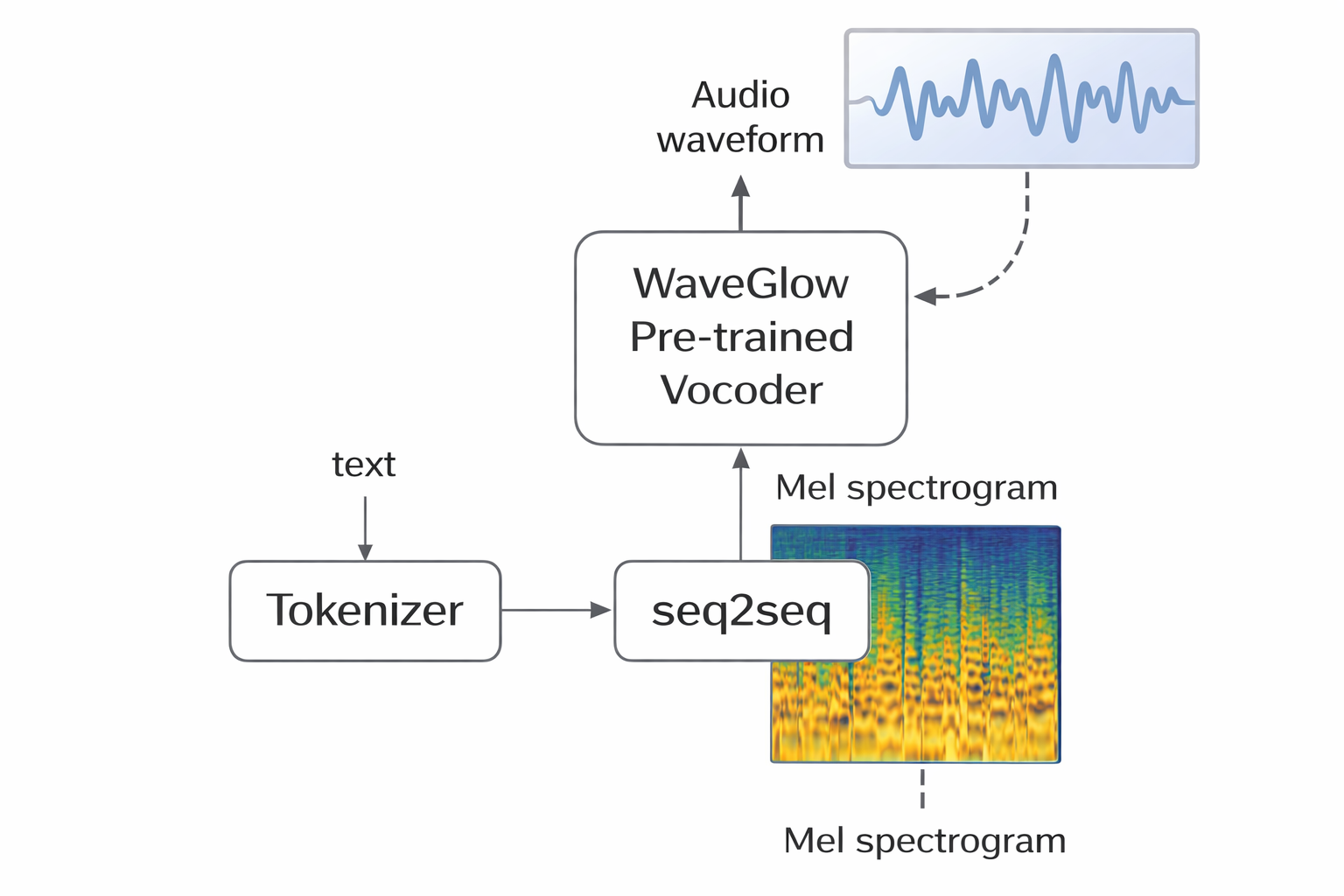}
    \caption{High level block diagram of \ftts{} architecture. We train the seq2seq block while using a pre-trained vocoder. 
    % The tokenizer is a standard character tokenizer.
    }
    \label{fig:tts_arch}
\end{figure}

% which must be possible if the model is a “purely data driven” system.
 \artts{} \cite{shen2018natural} is an \ftts{} deep learning architecture that consists of two main components. As seen in Figure \ref{fig:tts_arch}, the first component is a recurrent sequence-to-sequence (seq2seq) model with an attention mechanism, and a \WaveNet-based vocoder. The seq2seq model takes in a sequence of characters (text) and outputs a sequence of %well known
 mel-spectrogram features. These output mel-spectrogram features  pass through a pre-trained \WaveNet{} vocoder to generate the audio waveform (speech).

Specifically, the text to mel-spectrogram seq2seq model 
%is responsible for converting the input text into a sequence of mel-spectrogram features. It 
consists of an encoder and a decoder, %both of which are 
implemented as recurrent neural networks (RNNs). The encoder takes in the input text and encodes it into a sequence of high-level representations, which are then passed to the decoder. The decoder uses an attention mechanism to focus on different parts of the input sequence at each time step, allowing it to generate a sequence of mel-spectrogram features that %accurately 
represent the input text.
On the other hand, 
the \WaveNet \cite{prenger2019waveglow} vocoder is a deep generative flow-based model %.
% that uses autoregressive prediction to generate the audio waveform one sample at a time.
% The \WaveNet{} vocoder
% It, generally is 
trained on a large corpus of speech data, allowing it to generate highly realistic and natural-sounding speech.
The 
two stage 
architecture of seq2seq followed by vocoder (see Fig. \ref{fig:tts_arch}) allows generation of  high-quality and natural-sounding speech from input text. %making it a powerful tool for text-to-speech applications.

While using a pre-trained vocoder \cite{waveglow} 
%\footnote{https://github.com/NVIDIA/waveglow/} 
we built several \ftts{} systems by training several different seq2seq models as shown in Fig. \ref{fig:tts_arch}. The first seq2seq model was trained using the standard text, denoted by $w$ and mel-spectrogram extracted from $s(t)$ for the system denoted by \ftts{}.  Namely,
\[ 
\text \longrightarrow \fbox{\ftts{}} \longrightarrow \speech
\]
%This is a  corpora that is commonly used in a typical TTS system;  and 
The second system, called, \rtts{} while using the same pre-trained vocoder, used the reverse text\footnote{``dlrow olleh'' is the reverse text corresponding to ``hello world''} denoted by $\rw$ and mel-spectrogram extracted from reverse speech denoted by $\rs$ to train the seq2seq model. The trained \rtts{} is denoted by 
\[
\rtext \longrightarrow \fbox{\rtts} \longrightarrow \rspeech
\]
Note that $\rspeech$ has to be reversed for any further analysis, namely $\rrspeech$ (reverse of $\rspeech$). 
In both the  cases we use the same standard seq2seq architecture to train for the text to its corresponding mel-spectrogram, while using a pre-trained \WaveNet{} model. 
% Note that $\rs$ is synthetically constructed from $s(t)$ and not naturally produced by a human.

\section{Experimental Details}
\label{sec:experiments}
% \subsection{Dataset}
The \artts{} \cite{shen2018natural} model %\footnote{It is to be noted that we could have used any seq2seq model. We had this model running in our environment.} 
is trained on the text-speech paired data of English male speaker, released as part of the LIMMITS'25 grand challenge \cite{limmits25}.
% \footnote%{https://sites.google.com/view/limmits25}. 
The dataset consist of approximately $40$ hours of speech data spoken by a single speaker recorded at $48$ kHz sampling rate. Note that we do not experiment with different datasets or multiple number of speakers because that is not the focus of this experimental study.

The following data-prepossessing was done on the dataset to train the seq2seq model. The speech was
%is done on the speech and the text samples 1) 
down-sampled to $22.05$ kHz using {\tt sox} utility and 
%2) extracting the 
mel-spectrogram was extracted from the speech with window length of $1024$ samples, %$46.44$ msec, 
 hop length of $256$ samples %$11.6$ msec 
and $80$ mel channels. The text was converted to lower case, followed by normalization (numerical converted to words, for example ``10'' was converted to ``ten'', etc). We followed the same pre-processing for training the seq2seq model for different \ftts{} models. %both 
% \ftts{}, \rtts{}, BPE-\ftts{}, and BPE-\rtts{}.
% 3) the text is converted into a numeric sequence where every unique character of the text is assigned a unique number.

% \subsection{Model configuration}
% In this experiment, two models based on the tacotron-2\cite{shen2018natural} architecture were trained. One model was trained on the normal text and mel-spectrogram pair while the second model was trained on the pair of reversed text and reversed mel-spectrogram (reversed with respect to time-axis).

% The tacotron-2\cite{shen2018natural} module learns only to map the text sequence to mel-spectrogram. So, to convert the mel-spectrogram to audio, WaveGlow\cite{prenger2019waveglow} vocoder was used. Small experiments were conducted to verify that the WaveGlow vocoder was independent of the speaker and also independent of the orientation (normal or reversed with respect to the time axis) of the mel-spectrogram. It was confirmed that the vocoder could generate audio from reversed mel-spectrograms.

Comparing %Looking specifically at  
\ftts{} and \rtts{}, both %e2e-TTS and \rtts{} 
seq2seq models were trained under identical conditions, namely, on the same machine, using a batch size of $16$ and a learning rate of $2.5 *10^{-4}$. %$0.00025$. Both 
All the models were trained for $50000$ iteration. 
% Figure \ref{fig:loss_plot} shows the  plot of the loss for both the training set and the validation set for both  e2e-TTS and \rtts{} as a function of the number of iterations.

% Parth to put the figure associates with training loss here ...

\begin{figure}[!htb]
    \centering
     \begin{subfigure}{0.475\linewidth}
        \centering
        \includegraphics[width=1\linewidth]{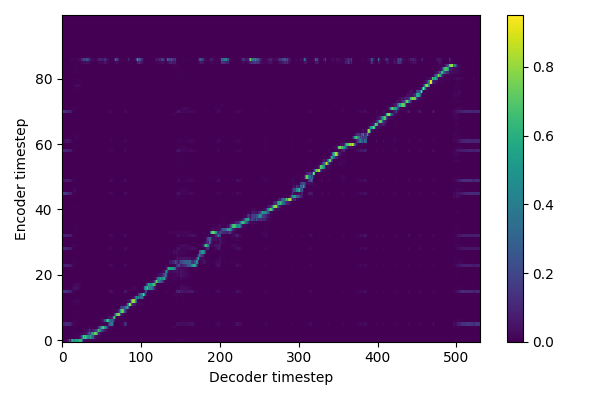} % Replace with your image file
        \caption{Alignment plot (\ftts{}).}
        \label{fig:image1}
    \end{subfigure}
    % 
    % \vspace{0.5cm} % Adjust space between images if needed
% 
    \begin{subfigure}{0.475\linewidth}
        \centering
        \includegraphics[width=1\linewidth]{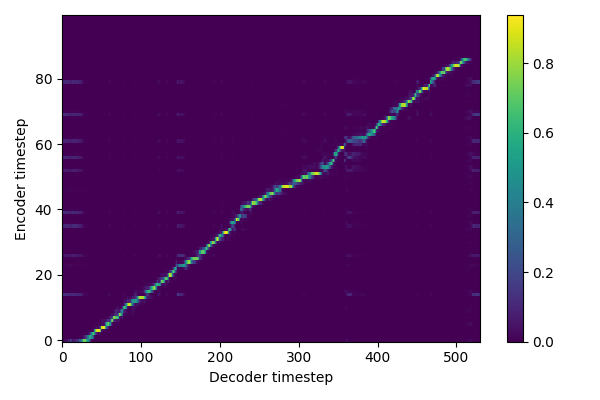} % Replace with your image file
        \caption{Alignment plot (\rtts{}).}
        \label{fig:image2}
    \end{subfigure}
    \caption{Alignment plots for both \ftts{} and \rtts{} 
    %The %training loss 
    %do not 
    show no observable differences.}
    \label{fig:alignment_plot}
\end{figure}

Figure \ref{fig:alignment_plot} shows the alignment plots between encoder and decoder steps generated during the training process at the $45000^{th}$ iteration. It can be observed that there is no \textit{noticable} differences in the two alignment plots, demonstrating that both the trained models \ftts{} (\rtts{}) effectively learned to convert text sequence, $\text$ ($\rtext$) into audio $\speech$ ($\rspeech$).

For the purposes of quantitative comparison, we selected $2500$ text samples, not used during seq2seq training, at random, from Librispeech-960 \cite{panayotov2015librispeech} corpus. We discarded all sentences that had less than $5$ words or  greater that $40$ words which resulted in a total of $1715$ sentences.  We generated audio, using the trained \ftts{} and \rtts{} models separately for these $1715$ sentences. 
A well documented problem in seq2seq model is its inability to identify the End-of-Sentence (\eos). 
A total of $14$ generated audio (see Table \ref{tab:stats}) by either \ftts{} ($10$) or \rtts{} ($4$) resulted in an extended duration (limited by the maximum duration of the audio chosen as the hyper-parameter) audio 
because they resulted in the seq2seq model not being able to identify \eos. 
We discarded these $14$ audio files from further analysis. All the analysis is based on the remaining $1701$ generated audio, common to both \ftts{} and \rtts{}.
 
\begin{table}[!htb]
    \centering
    \caption{Synthesized audio seq2seq \eos\ problem.}
    \label{tab:stats}
    \begin{tabular}{|c|c|c|} \hline
    Model     &  Generated (\#)& \eos\ issues (\#)\\ \hline
    \ftts{}     &  1715 & 10\\
    \rtts{}     &  1715& 4\\ \hline
    \end{tabular}
\end{table}

We evaluated the audio generated by \ftts{} and \rtts{} in two different ways. In the first set of evaluation,
we transcribed the generated audio using OpenAI's Whisper {\tt small.en} and WhisperX {\tt small} model \cite{radford2023robust}. Note that while we passed $\speech$ generated by \ftts{} 
% (BPE-\ftts{}) 
directly, we reversed the output of \rtts{} %, (BPE-\rtts{}), 
namely used $\rrspeech$ for transcription.  
As an objective evaluation metric, we computed the Word Error Rate (WER) and Character Error Rate (CER), by finding the edit distance between the transcribed text and the original text used for generating the audio. 
%Before computing the WER and CER, both the original text and the transcription text were converted into lowercase and all the punctuation's including commas and full-stops were removed from the text.  
Table \ref{table:wer_cer} captures the fidelity of the audio generated by \ftts{} and \rtts{}, %BPE-\ftts{} and BPE-\rtts{} 
in terms of %the proxy measure of 
WER and CER as a proxy measure. Note that smaller values of WER and CER are better and suggest that the generated audio is more reliably transcribed by the speech recognition engine.

% We evaluated the audio generated by \ftts{} and \rtts{} in two different ways. In the first set of evaluation, we selected $2000$ text samples, at random, from Librispeech-960 \cite{panayotov2015librispeech} corpus to generate audio waveform. We transcribed the generated audio using OpenAI's Whisper {\tt medium.en} model \cite{radford2023robust}. Note that while we passed $\speech$ generated by \ftts{} directly, we reversed the output of \rtts{}, namely used $\rrspeech$ for transcription.  As an objective evaluation metric, we computed the Word Error Rate (WER) and Character Error Rate (CER), by finding the edit distance between the transcribed text and the original text used for generating the audio. Table \ref{table:wer_cer} captures the performance of \ftts{} and \rtts{} in terms of these WER and CER. Note that smaller values of WER and CER are better.

%were computed by comparing the original text with the transcriptions of the generated speech samples, serving . The transcription was performed using 

% Both models successfully learned the mapping from text sequences to speech and were able to generate intelligible speech. For the evaluation of these models, 2000 random text samples were taken from {} and speech samples were generated. Word Error Rates (WER) and Character Error Rates (CER) were calculated between the original text and the transcriptions of the generated speech samples as objective evaluation metrics. For transcription, the "medium.en" model from OpenAI's Whisper\cite{radford2023robust} was used. 

Clearly, the speech synthesized by \rtts{} is better recognised by the ASR (both Whisper and WhisperX) than the conventional \ftts{} generated audio. It can be observed that the \rtts{} performance is $6.5\%$ ($5.6\%$) better than \ftts{} in absolute terms and $37.58\%$ ($34.36\%$) in relative terms for Whisper (WhisperX). It can also be observed that WhisperX, the more recent model performs, as expected, better than Whisper in all instances.

\begin{table}[!htb]
\centering
\caption{WER and CER (lower values better) calculated on $1701$ text samples, using \artts{} architecture.}
\label{table:wer_cer}
\begin{tabular}{|c|c|c|c|}
\hline\hline
% \multicolumn{2}{|c|}{}  & \multicolumn{2}{c|}{\artts{}} \\ \hline
& Model     & WER (\%)      & CER (\%)     \\ \hline\hline
\multirow{2}{*}{Whisper ({\tt small.en})} & \ftts{}       & 17.3         & 7.5         \\ \cline{2-4}
& \rtts{}     & \textbf{10.8} & \textbf{4.0} \\ \hline\hline
\multirow{2}{*}{WhisperX ({\tt small})} & \ftts{}       & 16.3         & 6.9         \\ \cline{2-4}
& \rtts{}     & \textbf{10.7} & \textbf{3.9} \\ \hline\hline
\end{tabular}
\end{table}

These observations indicate that %both  
\rtts{} %and BPE-\rtts{} are 
is
not only capable of learning the seq2seq mapping between the reverse text sequence and the corresponding mel-spectrogram of the reverse audio, but 
also generates speech that is \textit{better} transcribed by the automatic speech recognition engine; both Whisper \cite{openai_whisper} and WhisperX \cite{bain2023whisperxtimeaccuratespeechtranscription}.

% can also learn the mapping of the the compressed representation (BPE) of reverse text to its corresponding mel-spectrogram. In addition, the %r- systems, namely,  
% \rtts{} and BPE-\rtts{} are able to %also 
% generate audio waveform which are \textit{better} transcribed by the automatic speech recognition engine \cite{openai_whisper}.

%using reversed text and speech. 

% These results indicate that \rtts{} 
% %end-to-end systems are 
% is not only capable of learning the mapping between reverse text %character 
% sequences and the corresponding mel-spectrogram of the reverse audio, but generates audio waveform which are \textit{better} transcribed by automatic speech recognition engine \cite{openai_whisper}.
% %using reversed text and speech. 
% This suggests that (a) e2e TTS system (in our experiment, Tacotron-2) is primarily adopting a blind data-mapping exercise and it is not clear why the performance of speech recognition is better for audio generated using \rtts{}. 

% Additionally, the model trained on the reversed data seems to perform better than the model trained on the normal data

The second set of evaluations are based on human perception, in terms of mean opinion scores (MOS).  We asked $5$ subjects ($3$ male, $2$ female) to listen to a total of $16$ generated audio waveform ($8$ each generated using \ftts{} and \rtts{}) and give a score between $1$ and $5$  for both naturalness and intelligibility. The subjects were provided the information mentioned in Table \ref{tab:ni_index} as reference.
\begin{table*}[!htb]
    \centering
    \caption{Reference provided to subjects to mark naturalness and intelligibility.}% of the audio.}
    \label{tab:ni_index}
    \begin{tabular}{|c|p{3.5cm}|p{4.5cm}|} \hline\hline
     Score & Naturalness & Intelligibility \\ \hline\hline
      1   & awkward, robotic & cannot understand speech\\ \hline
      2   & strange/ not smooth & understandable; mostly unclear\\ \hline
      3   & like human voice & understandable; hard to follow\\ \hline
      4   & smooth and natural & easy to understand\\ \hline
      5   & {sounds} very natural & completely understandable \\ \hline\hline
    \end{tabular}
\end{table*}
Note that the subjects were not aware of which audio was generated by which system. Table \ref{table:MOS} shows the results of this perceptual evaluation in terms of MOS. While the MOS for both \ftts{} and \rtts{} are close to each other, the audio generated by \rtts{} gets a higher(better) MOS for both naturalness ($3.75$ in comparison to $3.4$) and intelligibility ($3.58$ compared to $3.38$). Again demonstrating the superior performance of  \rtts{} over \ftts{}.

\begin{table}[!htb]
\centering
\caption{Mean Opinion Scores (Naturalness and Intelligibility).} %of the generated audio.}
\label{table:MOS}
\begin{tabular}{|c|c|c|} 
\hline\hline
         & \ftts{} & \rtts{}  \\ 
\hline\hline
Naturalness & $3.40\pm0.11$         & \textbf{3.75 $\pm$ 0.22}          \\ \hline
Intelligibility & $3.38\pm0.27$          & \textbf{3.58 $\pm$ 0.60}           \\\hline\hline
\end{tabular}
\end{table}

In another perceptual evaluation test, the same $5$ subjects were presented with $5$ pairs of audio samples, generated using the same text ($\text$ for \ftts{} and $\rtext$ for \rtts{}), while one was generated using \ftts{} the other paired audio was generated using \rtts{}. These audio waveform were presented in random order, meaning in some instances the first audio waveform in the pair was from \ftts{} while in some instances the first waveform was from \rtts{}. The subjects were then asked to mark the audio sample which they felt was comparatively better of the two. Note that no specific instruction was communicated to the subjects in terms of what they should look for in the audio. In all, we got $5$ responses from each of the $5$ subjects, totaling $25$ responses as shown in Table \ref{tab:pair_test}. It can be observed that $92\%$ of the time ($23/25$) the subjects felt that the audio  generated by \rtts{} was better than the corresponding audio generated by \ftts{} and only in $2$ instances did the subjects choose the audio generated by \ftts{} over \rtts{} (see Table \ref{tab:pair_test}).
\begin{table}[!htbp]
    \centering
    \caption{Paired audio perceptual test. A $1$ represents the instance when \rtts{} generated audio was chosen as being better.}
    \label{tab:pair_test}
    % \resizebox{\linewidth}{!}
    {
    \begin{tabular}{|c|c|c|c|c|c|} \hline\hline
    & \subject{1} & \subject{2} & \subject{3} & \subject{4} & \subject{5} \\ \hline\hline
     \pair{1}    &  $0$& $1$& $1$& $1$& $1$\\
     \pair{2}    &  $1$& $1$& $1$& $1$& $1$\\
     \pair{3}    &  $1$& $0$& $1$& $1$& $1$\\
     \pair{4}    &  $1$& $1$& $1$& $1$& $1 $\\
     \pair{5}    &  $1$& $1$& $1$& $1$& $1$\\ \hline\hline
    \end{tabular}
}
\end{table}
Clearly, the perceptual evaluation, though by a small set of $5$ independent subjects, demonstrates the trend towards their preference for audio generated by \rtts{}. 

\begin{figure}[!htb]
    \centering 
\includegraphics[width=0.8\linewidth]{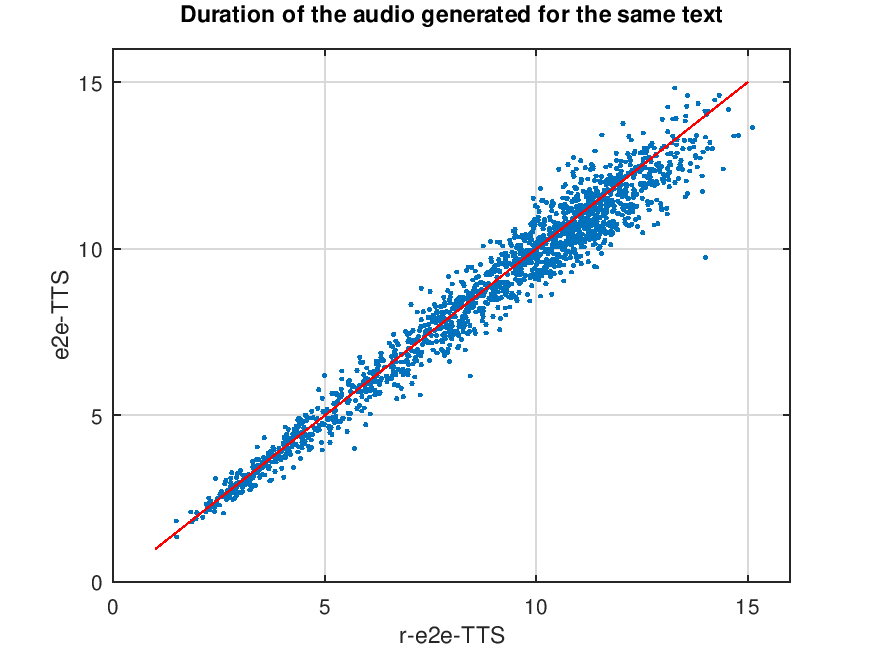}
    \caption{Scatter plot of duration of the audio for the same text produced by \ftts{} (y-axis) and \rtts{} (x-axis). The duration of audio generated by \ftts{} is overall longer than the audio generated by \rtts{}.}
    \label{fig:length_comparision}
\end{figure}

Further, it was observed that the total duration of the $1701$ audio generated by \rtts{} was $1.85\%$ shorter ($14786.21$ seconds compared to $15064.73$ seconds by \ftts{}) than that generated by \ftts{}.  On an average for the same $\text$ ($\rtext$ for \rtts{}) the \ftts{} generated  
$\speech$ was of longer duration ($8.85$ s) compared to the duration of $\rspeech$ ($8.69$ s).
Figure \ref{fig:length_comparision} captures the scatter plot of the duration of the audio produced by \ftts{} and \rtts{} for the same text. It can be observed that the number of points below the red diagonal far exceed the number of points above the diagonal. 

We conclude that the \rtts{} system seems to perform better in terms of lower WER and higher MOS (perceptual score) compared to \ftts{} everything else remaining same in terms of the architecture (\artts) and the hyper parameters, number of training samples, number of iterations.

\subsection{Effect of Text Tokenization}
Motivated by \cite{shah2025hamaraawaz}, in addition to \ftts{} and \rtts{} we also trained a set of models where the input text was encoded using byte pair encoding (BPE) \cite{sennrich-bpe} instead of directly using the characters in the text (see Fig. \ref{fig:tts_arch}).   For the same text input, generally BPE encoding results in lesser number of tokens, compared to character tokenization. As seen in Table \ref{tab:char_bpe_tokens} BPE results in $21$ tokens compared to $34$ character tokens for the example text \texttt{``The stars twinkle and shine bright''}.
\begin{table*}[!htb]
    \centering
    \caption{Tokenization for the sentence \texttt{``The stars twinkle and shine bright''}. %\cite{shah2025hamaraawaz}. 
    We use character tokenizer for \ftts{} and \rtts{} while we use Byte Pair Encoding (BPE) for BPE-\ftts{} and BPE-\rtts{}.}
    \label{tab:char_bpe_tokens}
    \begin{tabular}{p{4cm}|p{7cm}}\hline
    Tokenizer (\# of tokens) & Tokens \\ \hline
        character (34) & \texttt{t h e  \_ s t a r s \_  t w i n k l e \_  a n d  \_ s h i n e  \_  b r i g h t} \\ \hline
        BPE (21) & \texttt{the, \_, st, ar, s, \_t, w, in, k, le, \_, and, \_s, h, in, e, \_b, ri, g, h, t} \\ \hline
        BPE (token length $\leq2$) (24) & \texttt{\_t, he, \_s, t, ar, s, \_t, w, in, k, l, e, \_a, nd, \_s, h, in, e, \_b, r, i, g, h, t} \\ \hline
    \end{tabular}
\end{table*}
Tokenization algorithms, broadly work on the principle of iteratively merging frequently occurring pairs of characters or \sub-words\ to create a vocabulary of tokens that can represent the language's vocabulary efficiently. This as we will observe later, results in better machine transcription of the generated  audio. We call these TTS systems, 
% \[ 
% \text \longrightarrow \overbrace{\fbox{BPE} \longrightarrow \fbox{\ftts{}}}^{\mbox{BPE-\ftts{}}} \longrightarrow \speech
% \]
% and
% % We evaluate the audio output \ftts{}(
%  \[
% \rtext \longrightarrow \overbrace{\fbox{BPE} \longrightarrow \fbox{\rtts{}}}^{\mbox{BPE-\rtts{}}} \longrightarrow \rspeech.
% \]

\[ 
\text \longrightarrow \overbrace{\fbox{BPE} \rightarrow \fbox{\lengthregulator} \rightarrow \fbox{\ftts{}}}^{\mbox{BPE-\ftts{}}} \longrightarrow \speech
\]
and
% We evaluate the audio output \ftts{}(
 \[
\rtext \longrightarrow \overbrace{\fbox{BPE} \rightarrow \fbox{\lengthregulator} \rightarrow \fbox{\rtts{}}}^{\mbox{BPE-\rtts{}}} \longrightarrow \rspeech.
\]
Where \lengthregulator\ is a length-regulator operator which modifies the  length of the BPE tokenized sequence. The operator \lengthregulator\ expands the sequence by repeating each token to reflect its length. As an example,
\[
\texttt{bright} \rightarrow \fbox{BPE} \rightarrow \lbrack \texttt{br, i, ght}\rbrack \rightarrow \fbox{\lengthregulator} \rightarrow \lbrack \texttt{br,  br, i, ght, ght, ght} \rbrack
\]
the token ``{\tt br}'' of length two occurs twice and the token ``{\tt ght}'' of length three occurs thrice while the token `{\tt i}' of length 1 occurs just once. We find that \lengthregulator\ is crucial to produce intelligible speech when using \artts.

\begin{figure}[!ht]
    \centering
     \begin{subfigure}{0.65\linewidth}
        \centering
        \includegraphics[width=1\linewidth]{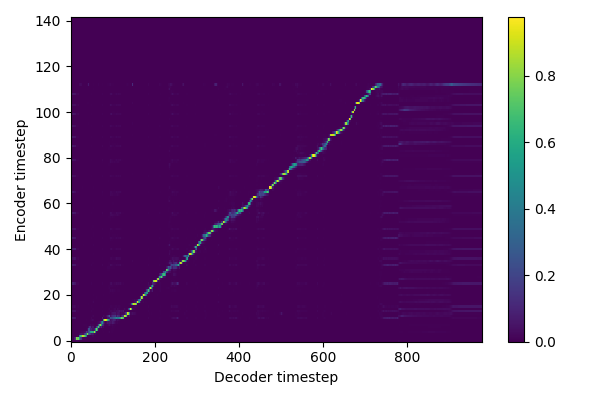} % Replace with your image file
        \caption{BPE-\ftts{}}
        \label{fig:bpe_image1}
    \end{subfigure}
    % 
    % \vspace{0.5cm} % Adjust space between images if needed
% 
    \begin{subfigure}{0.65\linewidth}
        \centering
        \includegraphics[width=1\linewidth]{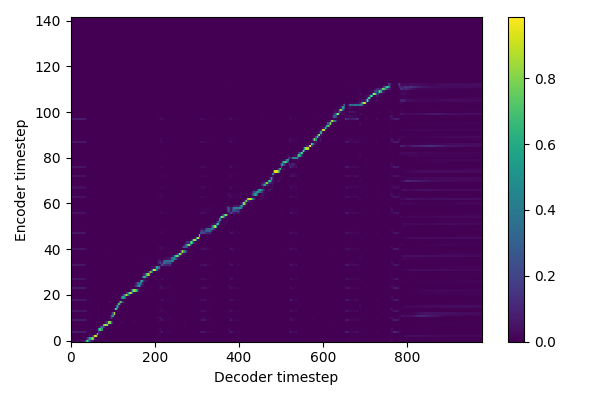} % Replace with your image file
        \caption{BPE-\rtts{}}
        \label{fig:bpe_image2}
    \end{subfigure}
    \caption{Alignment plots for both BPE-\ftts{} and BPE-\rtts{} 
    %The %training loss 
    %do not 
    show no observable differences.}
    \label{fig:bpe_alignment_plot}
\end{figure}

As mentioned earlier, the BPE models were trained on $\text$ for BPE-\ftts{} while it was trained with $\rtext$ for BPE-\rtts{}.
Figure \ref{fig:bpe_alignment_plot} shows the alignment plots between the encoder and decoder steps. From the plots it can seen clearly that the trained BPE models BPE-\ftts{} (BPE-\rtts{}) are able to effectively convert the BPE sequence to speech.
As earlier
% Similarly, 
identical training condition was enabled for training both BPE-\ftts{} and BPE-\rtts{}.  We trained the \SP\ \cite{kudo2018sentencepiece}, 
    with BPE 
    {(--vocab\_size=n, --model\_type={\tt bpe}, --split\_by\_whitespace=False)}. 
We use $n=50$ and impose a constraint on the token length to be $2$. 

These BPE models were also evaluated on the same set of $1715$ sentences. We generated separate audio files for all the $1715$ files for both the models. A total of $7$ (see Table \ref{tab:bpe_stats}) generated audio files  suffered from \eos\ problem, while BPE-\ftts{} resulted in $4$ instances of \eos\ BPE-\rtts{} resulted in $3$. So during the evaluation these $7$ files were discarded and the evaluation process was performed on the remaining $1708$ files.

\begin{table}[!htb]
    \centering
    \caption{Synthesized audio seq2seq \eos\ problem using BPE-\ftts{} and BPE-\rtts{}.}
    \label{tab:bpe_stats}
    \begin{tabular}{|c|c|c|} \hline
    Model     &  Generated (\#)& \eos\ issues (\#)\\ \hline
    BPE-\ftts{}     &  1715 & 4\\
    BPE-\rtts{}     &  1715& 3\\ \hline
    \end{tabular}
\end{table}

Along earlier lines, we computed the WER and CER of the audio files generated using BPE models and used the same ASR models for transcription.
%we used OpenAI's Whisper {\tt small} and WhisperX's {\tt small} models. 
Table \ref{tab:bpe_wer_cer} shows that though BPE-\rtts{} ($9.4\%$) performs better than BPE-\ftts{} ($9.6\%$), the improvement,  $0.2\%$ ($0.4\%$) in absolute terms and $2.1\%$ ($4.2\%$) relative terms for Whisper (WhisperX) is significantly dwarfed compared to using text without tokenization, namely, \ftts{} and \rtts{}.

Note that %Further, it can be observed that 
the 
audio produced by BPE models (both BPE-\ftts{} and BPE-\rtts{}; Table \ref{tab:bpe_wer_cer}) result in better WER ($9.6\%$) than the best when the input text is not encoded using BPE  ($10.7\%$; Table \ref{table:wer_cer}).

\begin{table*}[!htb]
\centering
\caption{WER and CER (lower values better) calculated on $1708$ text samples on audio generated by BPE-\ftts{} and BPE-\rtts{}.}
\label{tab:bpe_wer_cer}

\begin{tabular}{|c|c|c|c|}
\hline\hline
& Model     & WER (\%)      & CER (\%)      \\ \hline\hline
\multirow{2}{*}{Whisper ({\tt small.en})} & BPE-\ftts{}       & 9.6         & 3.4          \\ \cline{2-4}
& BPE-\rtts{}     & \textbf{9.4} & \textbf{3.3} \\ \hline\hline
\multirow{2}{*}{WhisperX ({\tt small})} & BPE-\ftts{}       & 9.5         & 3.4          \\ \cline{2-4}
& BPE-\rtts{}     & \textbf{9.1} & \textbf{3.2} \\ \hline\hline
% BPE-\ftts{}   & 7.09          & 2.94          \\ \hline
% BPE-\rtts{} & \textbf{6.46} & \textbf{2.44} \\ \hline\hline
\end{tabular}
\end{table*}

\subsection{Architecture Dependency}

To validate if the performance of \rtts{} being better than \ftts{} was dependent on the auto-regressive nature of  \artts, we experimented with 
% \red
{
%Additional, we did some experiments with 
\vits{} (Variational Inference Text-to-Speech) \cite{kim2021conditional} a non-autoregressive TTS architecture. Appendix \ref{appendix:2}  
% This model was trained in \ftts{} and \rtts{} configurations. 
% Table \ref{tab:tacotron2vsvits} 
captures the salient differences between the \artts{} model and \vits{} model.
}
\begin{table}[!htb]
\centering
\caption{WER and CER (lower values better) calculated on {{$1715$}} text samples using \vits{} architecture.}
\label{table:wer_cer_vits}
\begin{tabular}{|c|c|c|c|}
\hline\hline
% \multicolumn{2}{|c|}{}  & \multicolumn{2}{c|}{\vits{}} \\ \hline
& Model     & WER (\%) & CER (\%) \\ \hline\hline
\multirow{2}{*}{Whisper ({\tt small.en})} & \ftts{}       & 12.36 & 6.37 \\ \cline{2-4}
& \rtts{}     & \textbf{11.80} & \textbf{5.90} \\ \hline\hline
\multirow{2}{*}{WhisperX ({\tt small})} & \ftts{}       & 12.11 & 6.18 \\ \cline{2-4}
& \rtts{}     & \textbf{11.69} & \textbf{5.80} \\ \hline\hline
\end{tabular}
\end{table}
% \red

{
Table \ref{table:wer_cer_vits} shows the objective evaluation results of the \vits{} model.
% From the table it can be seen that even for \vits{} architecture the model performs in a similar way that the Tacotron-2 performed.
The speech synthesized by \rtts,  a model trained using $\rtext$ and $\rspeech$ %with \vits{} architecture 
is better transcribed by 
%both Whisper and WhisperX 
the ASR compared to the \ftts{} (trained using $\text$ and $\speech$) for the same \vits{} architecture. As seen in Table \ref{table:wer_cer_vits} the performance of \rtts{} is $0.56\%$ ($0.47\%$) better than \ftts{} in absolute terms 
%and $4.53\%$ ($3.47\%$) in relative terms 
for Whisper (WhisperX). We can conclude that %It can be observed that 
the performance of the \rtts{} is better that the performance of \ftts{} \textit{irrespective} of the TTS architecture, as seen in Table \ref{table:wer_cer} for \artts{} and Table \ref{table:wer_cer_vits} for \vits{}. 
}

{
Additionally, on an average for the same text $\text$ ($\rtext$ for \rtts{}) the duration of the generated audio was $8.62$ s for \ftts{} compared to a duration of $8.54$ s for the \rtts{} system (shorter by $0.93\%$). A trend similar to that observed using the \artts{} architecture.
}

\subsection{Reverse Text and Forward Speech (\RevTxtFwdSpeech)}
\label{sec:additional_experiments}

Encouraged by the superior performance of \rtts{} independent of the architecture (autoregressive or non-autoregressive) we wanted explore if training with reverse text ($\rtext$) and normal speech (\speech), namely, \RevTxtFwdSpeech{} was trainable.
We trained an \ftts{} systems using both \artts{} and \vits{} architectures. 
\begin{figure}[!htb]
    \centering
     \begin{subfigure}{0.65\linewidth}
        \centering
        \includegraphics[width=1\linewidth]{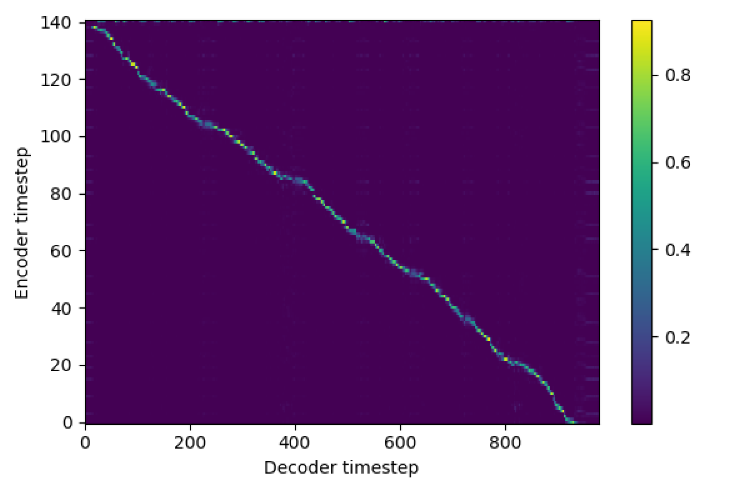} % Replace with your image file
        \caption{Alignment plot \RevTxtFwdSpeech\ (\artts)}
        \label{fig:rtfs-tacatron}
    \end{subfigure}
    % 
    % \vspace{0.5cm} % Adjust space between images if needed
% 
    \begin{subfigure}{0.65\linewidth}
        \centering
        \includegraphics[width=1\linewidth]{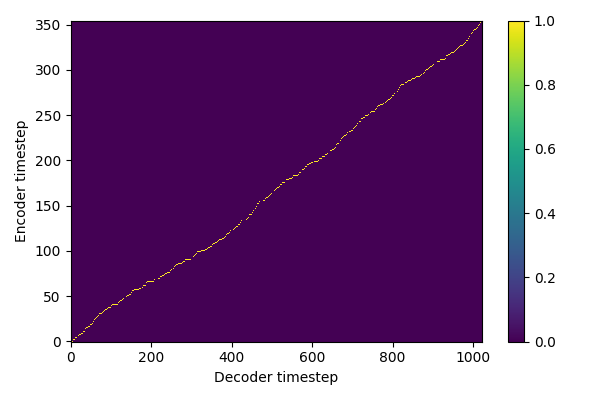} % Replace with your image file
         \caption{Alignment plot \RevTxtFwdSpeech\ (\vits)}
        \label{fig:rtfs-vits}
    \end{subfigure}
    \caption{Alignment plots for training \RevTxtFwdSpeech. (a) \artts{}, (b) \vits.}
    \label{fig:alignment_plot_RTFS}
\end{figure}
Figure \ref{fig:rtfs-tacatron} shows the alignment plot between the reversed text and normal speech frames generated by the \artts{} model. It can be clearly observed that the model is able to \textit{identify} and \textit{map} the initial  text character tokens to the final audio frames. Notice that the slope of the alignment curve is diagonally opposite to the alignment plots in Fig. \ref{fig:alignment_plot} and Fig. \ref{fig:bpe_alignment_plot}.  As can be seen from Table \ref{tab:RTFS_wer_cer} the performance in terms of WER and CER on 1711 audio transcriptions (of the 1715 audio files; 4 samples suffered from the EoS problem) is better that what has been observed for \ftts{}, \rtts{} in Table \ref{table:wer_cer} as well as the performance of BPE-\ftts{} and BPE-\rtts{} as seen in Table \ref{tab:bpe_wer_cer}. 

% Additionally, it can be seen that the performance in terms of both WER and CER is close to that of the performance of the reverse models (r-e2e-TTS and r-BPE-e2e-TTS). The WER performance for Whisper (WhisperX) is 9$\%$ (8.31$\%$) while the CER performance is 3$\%$ (3.73$\%$).

While \artts{} model is able to learn, in contrast \vits{} architecture shows no such alignment when an \ftts{} as can be seen in Fig. \ref{fig:rtfs-vits}.
The reason for 
the 
differences in the learning behaviour can be attributed to 
the differences in the \textit{alignment strategies} used by \artts{} and \vits. 
While \artts{} adopts uses the Location-Sensitive-Attention mechanism which is an extension of Bahadanau attention \cite{bahdanau2014neural} to enable \textit{soft-alignment} between the text tokens and the speech frames, \vits{} uses \textit{Monotonic Alignment} Search (MAS)\cite{kim2020glow} to align the text and speech. 
MAS introduces a constraint which requires the input text-sequence and the mapped audio-sequence have a strict left-to-right (monotonic) correspondence which results in  necessarily mapping the first text token to the first audio frame in the generated sample. This MAS constraint results in the decoder in the TTS model not to \textit{attend} to any of the previous tokens which are aligned already resulting in no alignment between the text and the audio.
Note that the \textit{soft-alignment} gives  \artts{}   a free-hand to explore all the possible relations between the text-tokens and the audio-frames, the \textit{monotonic alignment} in \vits{} restricts the alignment between reverse input text and the forward audio.
\begin{table}[!htb]
    \caption{WER and CER ($\downarrow$ better) calculated on 1711 text samples on audio generated by \RevTxtFwdSpeech. Using \artts.}
    \label{tab:RTFS_wer_cer}
    \centering
    \begin{tabular}{|c|c|c|c|}
        \hline\hline
         & Model & WER(\%) & CER (\%) \\
         \hline\hline
        Whisper ({\tt{small.en}}) & \RevTxtFwdSpeech & 9 & 3 \\
        \hline\hline
        WhisperX ({\tt{small}}) &  \RevTxtFwdSpeech & 8.31 & 3.73 \\
        \hline\hline
    \end{tabular}

\end{table}
\begin{table*}[!htb]
% \caption{Summary of Results from all the experiments. \NA{} indicates the model does not train.}
\caption{Summary of Results from all the experiments.}
\label{tab:results_summary}
\centering
\begin{tblr}{
  cells = {c},
  % Header merges
  cell{1}{1} = {r=2}{},    % TTS Arch spans 2 header rows
  cell{1}{2} = {r=2}{},    % Text Embedding spans 2 header rows
  cell{1}{3} = {r=2}{},    % model type spans 2 header rows
  cell{1}{4} = {c=2}{},    % Whisper spans 2 columns
  cell{1}{6} = {c=2}{},    % WhisperX spans 2 columns
  % Row-group merges in first column (TTS Arch)
  cell{3}{1,2} = {r=3}{},
  cell{6}{1,2} = {r=3}{},
  cell{9}{1,2} = {r=2}{},
  vlines,
  % Full-width horizontal lines (now 7 columns)
  hline{1,3,6,9,11} = {-}{},
  % Line under grouped headers (Whisper / WhisperX), now cols 4-7
  hline{2} = {4-7}{},
  % Internal lines excluding the first column (because it's row-spanned),
  % now cols 2-7
  hline{4-5,7-8,10} = {3-7}{},
}
\textbf{TTS Arch} & \textbf{Text Embed} & \textbf{Model type} & \textbf{Whisper} &  & \textbf{WhisperX} &  \\
                  &                         &                     & WER              & CER & WER               & CER \\

{\textbf{\artts{}}} & \textbf{char} & \ftts{}         & 17.32 & 3.44 & 16.33 & 6.94 \\
                      &                    & \rtts{}       & 10.79 & 3.28 & 10.69 & 3.9  \\
                      &                    & \RevTxtFwdSpeech & \textbf{9}     & 3    & \textbf{8.31}  & 3.73 \\

{\textbf{\vits{}}}   & \textbf{char} & \ftts{}         & 12.36 & 6.37 & 12.11 & 6.18 \\
                      &                    & \rtts{}       & 11.83 & 5.96 & 11.69 & 5.8  \\
                    %   &                    & \RevTxtFwdSpeech & \NA    & \NA   & \NA    & \NA   \\
                      &                    & \RevTxtFwdSpeech & 101.69    & 85.03   & 110.15    & 84.55   \\

{\textbf{\artts{}}} & \textbf{BPE}       & BPE-\ftts{}     & 9.58  & 3.44 & 9.55  & 3.41 \\
                      &                    & BPE-\rtts{}   & 9.39  & 3.28 & 9.16  & 3.23
\end{tblr}
\end{table*}

Table \ref{tab:results_summary} summarizes all the experimental results. It can be observed that the best performance in terms of WER (a proxy measure to determine the intelligibility of synthesized speech) is enabled using \RevTxtFwdSpeech{} model with \artts{} architecture. This experimental exploration seems to suggest that \ftts{} systems are purely data driven and is not constrained by the Human Articulatory Constraints that restricts the certain sequence of phonemes in human generated speech (see Appendix \ref{appendix:1}).

\section{Conclusion}
\label{sec:conclusion}
We hypothesized that there might be hidden \textit{natural} properties in the speech produced by human because of the constraints imposed by the anatomy of the human articulatory mechanism and asked ourselves if these properties, if any, influenced the training of an \ftts{} system. %(specifically). 
Our experiments, based on \artts{} and \vits{} architecture
%\footnote{should hold true for any other e2e TTS architecture} as the e2e system, %
suggest that e2e deep learning systems  
execute as a \textit{brute-force} and \textit{blind} data-mapping exercise, seq2seq learning the text sequence and the corresponding mel-spectrogram of the audio.

All our experiments used read speech (it is common to use read speech in TTS literature), however,  it remains to be seen if the articulately transition constraints are more visible if 
spontaneous speech was used to train an \ftts{} system. In normal communication, humans speak spontaneously, which introduces pauses and variations in word arrangement within sentences randomly. It remains to be explored whether \ftts{} systems exploit these properties.

We enumerate some of our observations, many of  which, in our opinion, 
require %\red
{further exploration for deeper analysis and better understanding.}
\begin{enumerate}%[label=(\alph*)]

% \item One could attribute the deep 
% (seq2seq) model to learn the association between the text string and the acoustic (mel-spectrogram) purely as a data mapping exercise. Negating our hypothesis that e2e systems might be exploiting \textit{hidden} patterns due to constraints imposed by anatomy of articulators.
    \item %Tacotron-2 based 
    End-to-end TTS (\ftts) systems  seem to primarily adopt a blind data-mapping exercise and train well for forward-forward, reverse-reverse, forward-reverse \ftts{} systems, especially when \artts{} architecture is used. 
    \begin{enumerate}
    \item All the %four 
    models namely, \ftts, \rtts{} (both \artts{} and \vits), BPE-\ftts{} and BPE-\rtts{} are able to train and %are able to 
    produce natural and intelligible speech.
        \item It is not clear why the performance, in terms of speech recognition by ASR and speech perception by humans, is better for audio generated using \rtts{} and BPE-\rtts. A deeper study might shed some light on this aspect. 
    \end{enumerate}
    
\item %It is not clear, why the 
The audio generated by BPE-\rtts{}, \rtts{}, and \RevTxtFwdSpeech, are transcribed more accurately than the audio generated by \ftts{} by the state of the art ASR engines, Whisper and WhisperX. This needs to be investigated further.

\item The use of BPE results in lower (better) WER, demonstrating that use of a text encoder
to represent input text to train the seq2seq model seems to better capture the text-speech mapping. It is probably because longer length tokens can capture \textit{contextual information} embedded in the %mel-spectrogram of the 
speech better than character based tokens. 
\begin{enumerate}
    \item However, using longer size tokens ($\ge 3$) when using \artts{} architecture resulted in poor quality audio generation in terms of both naturalness and intelligibility. This requires further investigation as well.
\end{enumerate}

\item 
While there are no challenges in training an \rtts{} (BPE-\rtts), 
the appropriateness of the data used to train requires attention.
\begin{enumerate}
\item The $\rs$ (reverse audio) used for training  is \textit{artificial} and synthetically generating; the  reverse speech used in our experiments is not really equivalent to a \textit{human speaking  in reverse}.
\item Our attempt to get a few subjects (see Appendix \ref{appendix:3}) to speak a sentence in reverse showed that it was not only impossible to utter the sentence (expressed by all the subjects) without unusual pauses, but also when the recorded sentence was played in reverse, it sounded gibberish.
\end{enumerate}

\end{enumerate}

\bibliographystyle{splncs04}
\bibliography{mybib}

\appendix

\section{Articulatory Factors Underlying the Difficulty of Producing ``txet''}
\label{appendix:1}

Producing the non-word ``txet'' is articulatorily more demanding than producing the real word ``text'' due to differences in phoneme sequencing and motor coordination requirements. In ``text'' (/t$\epsilon$kst/), the phoneme sequence follows common English phonotactic patterns, progressing from an alveolar stop to a vowel, then to a velar stop, a fricative, and finally returning to an alveolar stop. This sequence, alveolar $\rightarrow$ vowel $\rightarrow$ velar $\rightarrow$ fricative $\rightarrow$ alveolar---is typical in English and allows for smooth and incremental transitions between articulatory configurations.

In contrast, ``txet'' (/tx$\epsilon$t/) imposes an unusual and less natural phoneme order. The transition from /t/ (an alveolar stop) directly to /x/ (a velar fricative) occurs without an intervening vowel. This forces a rapid and substantial backward movement of the tongue, shifting from an alveolar closure to a velar constriction within a very short temporal window. Such a transition is biomechanically demanding, as it requires precise timing and large-scale tongue repositioning without the articulatory ``buffer'' typically provided by a vowel.

These difficulties are further compounded by coarticulation effects. In natural speech production, articulatory gestures are planned anticipatorily, allowing upcoming sounds to influence current articulator positioning. In ``text'', the presence of a vowel facilitates gradual tongue repositioning and smooth coarticulatory overlap between consonants. In ``txet'', however, the absence of an intermediate vowel before /x/ disrupts this process, forcing an abrupt articulatory jump that limits coarticulatory smoothing.

As a result, producing ``txet'' demands greater articulatory precision, tighter temporal coordination, and increased motor control compared to ``text''. This elevated motor effort increases both cognitive and physical load during speech production, providing a plausible explanation for the increased difficulty associated with producing the non-word relative to the real word.

\section{\artts{} versus \vits}
\label{appendix:2}
\artts{} and \vits{} represent two distinct generations of text-to-speech architectures with fundamentally different design philosophies. \artts{}, introduced in 2018, follows a two-stage autoregressive pipeline in which text is first converted into a mel-spectrogram using an encoder–decoder architecture with location-sensitive attention, and raw audio is subsequently synthesized using a separate neural vocoder such as WaveGlow. This staged design results in frame-wise, sequential generation, which limits inference speed despite its strong alignment learning capabilities.

In contrast, \vits{}, proposed in 2021, adopts a single-stage, non-autoregressive framework that integrates text encoding, alignment learning, and waveform generation into a unified model. Its architecture comprises a prior encoder, posterior encoder, and decoder trained jointly within a normalizing flow framework, eliminating the need for an external vocoder. Alignment between text and speech is learned using Monotonic Alignment Search (MAS), enabling stable and interpretable duration modeling. Additionally, \vits{} incorporates a duration predictor and an adversarial discriminator, allowing fully parallel waveform generation and significantly improved inference efficiency.

Overall, while \artts{} relies on explicit attention mechanisms and a modular pipeline that trades speed for interpretability, \vits{} achieves faster and more scalable synthesis through end-to-end training, implicit alignment learning, and parallel generation, marking a substantial advancement in neural TTS system design. Details in Table \ref{tab:tacotron2vsvits}.

\begin{table*}[!htb]
\centering
\caption{Comparison between \artts{} and \vits{} models}
\label{tab:tacotron2vsvits}
\begin{tabular}{|c|p{0.4\linewidth}|p{0.4\linewidth}|}
\hline
{\textbf{Aspect}} & {\textbf{\artts{}} (2018)} & {\textbf{\vits{}} (2021)} \\ \hline
\multirow{1}{*}{\textbf{Architecture}} 
  & Two-stage (autoregressive)& Single-stage (non-autoregressive)\\
%   & Autoregressive pipeline & Non-autoregressive pipeline \\
\hline
\multirow{2}{*}{\textbf{Structure}} 
  & Encoder-decoder with attention (text to spectrogram) & Integrated  prior encoder, posterior encoder, decoder \\
  & WaveGlow vocoder for raw audio from spectrogram& Duration predictor + discriminator \\
\hline
\multirow{1}{*}{\textbf{Efficiency}} 
  & Slower (frame-wise generation) & Faster (parallel generation) \\
%   & (frame-wise generation) &(parallel generation) \\
\hline
\multirow{2}{*}{\textbf{Alignment}} 
  & Location-sensitive attention & Monotonic Alignment Search \\
  & — & Normalizing flow framework \\
\hline
\end{tabular}
\end{table*}

\section{Can $\rspeech$ be produced by Human?}
\label{appendix:3}
\begin{figure}[!htb]
    \centering
    \includegraphics[width=0.5\textwidth]{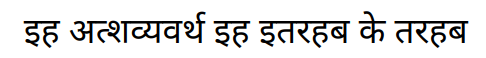}
    
    (a)
    
    \includegraphics[width=0.5\textwidth]{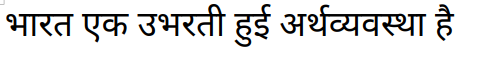}
    
    (b)
    \caption{(a) $\rs$ and (b) $\speech$}
    \label{fig:rf-hindi}
\end{figure}
We asked $5$ subjects who knew Hindi to {read} the Hindi sentence in Fig \ref{fig:rf-hindi} (a) which is the reverse of \ref{fig:rf-hindi} (b). Even after extensive rehearsing, all the subjects expressed difficulty in reading the reverse text. They claimed that the sentence had no semantics and there were not many known words. We further observed that the read reverse speech had no intonation, the sentence was made as Word by word utterance and had too many unrealistic pauses. Further, playing the recorded $\rs$ in reverse neither sounded natural nor was it intelligible.

We conclude that true reverse speech can not be produced by Human which makes us believe collecting \textit{actual} reverse speech data is very difficult.

\end{document}